# Contextual Mobile Learning Strongly Related to Industrial Activities: Principles and Case Study

Bertrand DAVID, Chuantao YIN and René CHALON
LIESP Laboratory, Ecole Centrale de Lyon, Ecully, France

*Abstract*—M-learning (mobile learning) can take various forms. We are interested in contextualized M-learning, i.e. the training related to the situation physically or logically localized. Contextualization and pervasivity are important aspects of our approach. We propose in particular MOCOCO principles (Mobility - COntextualisation - COoperation) using IMERA platform (Mobile Interaction in the Augmented Real Environment) covering our university campus in which we prototype and test our approach. We are studying various mobile learning contexts related to professional activities, in order to master appliances (Installation, Use, Breakdown diagnostic and Repairing). Contextualization, traceability and checking of execution of prescribed operations are based mainly on the use of RFID labels. Investigation of the appropriate training methods for this kind of learning situation, applying mainly a constructivist approach known as "Just-in-time learning", "learning by doing", "learning and doing", constitutes an important topic of this project. From an organizational point of view we are in perfect symbiosis with EPSS - Electronic Performance Support System [12] and our objective is to integrate learning in professional activities in three ways: 1/ before work i.e. to learn about coming actions, 2/ after work i.e. to learn about past actions to understand what happened and accumulate experience, 3/ during work i.e. to master the problem just-in-time We are also studying an appropriate relationship between the just-in-time M-learning approach and preliminary training performed in a serious game approach based on typical action scenarios.

## I. INTRODUCTION

As mobile technologies are becoming more and more widespread in people's daily life, there has been a huge increase in studies and experiments regarding the use of mobile technologies for learning and training in professional and industrial situations. More and more people are starting to benefit from various mobile learning technologies, but in the meanwhile arguments for the mobile learning definition have never stopped evolving over the years.

Compared to e-learning, we identified mobile learning cartography [1] in relation with ubiquitous computing, mobile computing and wearable computing, a mobile learning cartography as shown in Fig. 1.

Four essential characteristics are proposed to describe mobile learning situations [3]: devices, mobility, context, and location. Several categories of mobile learning can be identified according to the variation of these characteristics.

We consider it very important to separate M-learning into two categories in relation with the context. Either the learning activity is totally independent from the actor's location and the context in which he/ she is evolving, taking into account only the opportunity to use mobile devices to learn (in public transportation, waiting for the bus, etc.) or on the contrary, the learning activity relates to the location (physical, geographical or logical) of the actor and the context in which he/ she is evolving. We are mainly concerned with this second category of mobile learning, which is also called contextual mobile learning or context-aware mobile learning.

The learning context in mobile learning is a very important aspect, which can be described as "any information that can be used to characterize the situation of learning entities that is considered relevant to the interactions between a learner and an application" [4]. Three categories are considered: computing context, user context, and physical context [5]. Thanks to the development of wearable computers which are complemented by accessorial sensors, such as GPS receivers, RFID readers [6], cameras, etc, and software sensors, such as network congestion manager, web log analyzer, user behavior analyzer etc, the learning context such as learners' location, activity, network connectivity, learning situation, etc, can be captured to improve the learning activities.

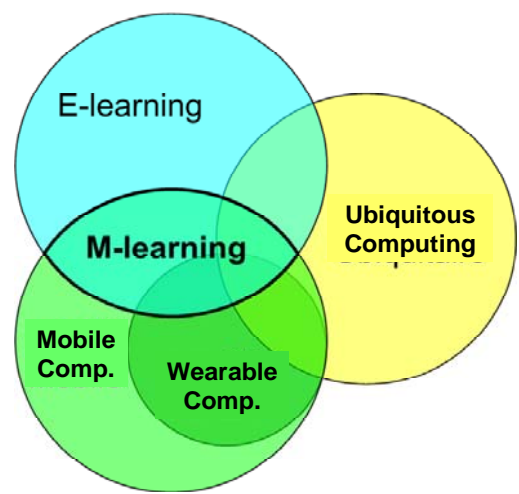

Figure 1. Mobile learning cartography

On the other hand, Mixed Reality [7] better known as Augmented Reality, appeared in 1993 [8], is becoming increasingly essential. This technology attempts to merge the physical and numerical worlds in order to facilitate the user's task with special devices and particular interaction techniques. A lot of hardware materials (like see-through goggles) and software applications were developed, which are very helpful for the mobile learning research in specific situations.

Study of learning methods is another important aspect of mobile learning. New and more appropriate methods are needed for the great revolution of mobile learning compared to the traditional classroom methods. We consider that these mobile learning methods should take into account the specific conditions in which learning occurs: limited time, particular working conditions and operational result orientation. Main learning method characteristics are such as problem-based learning [2], case-based learning [9], scenario-based learning [10], and just-in-time learning.

Several existing e-learning projects provide us with much useful inspiration for designing new mobile learning systems. What we are interested in is applying mobile learning theories to industrial learning and training activities. Our contextual mobile learning system is part of a larger project called Help Me to Do (HMTD) [11], which aims at assisting different users in domestic, public or professional mastering of appliances by wearable computer. The aim of this paper is to discuss new learning methods to be used in and for industry in relation with web-based knowledge access and digital era evolutions and their integration in M-learning MOCOCO environment to allow the user to learn just-in-time use, diagnosis and repair procedures.

In the following sections, we present a discussion of learning in and for industry. Next, augmentations of human, environment and appliance in order to facilitate coordination, cooperation and conversation are described, as well as global process and learning unit production. We finish by a case study and a conclusion

## II. LEARNING IN & FOR INDUSTRY

Learning of industrial professionals is an important challenge which can accommodate different solutions in respect with different professional situations. Without identifying all these situations, we can point out two, which are extreme and thus representative: before starting professional work, mainly during university studies and during professional, in the field, life. University studies form an appropriate period in which to assimilate important theories and generic methodological approaches. In the field, these aspects are more difficult to acquire. On the contrary, practical, precise behaviors, operations and gestures are difficult to acquire at university and easy in the field. It is important to take into account these two kinds of knowledge to be assimilated.

As an old Chinese proverb says "give a fish to a man and he can eat one day, give him a fishing rod and he can eat all his life". This proverb is a motto (device) for university studies, allowing students to understand generic theories and approaches on which they can build their life. But to cook a particular fish is also important. Unfortunately, university is not an appropriate place to learn cooking. Industry and in the field situations are more appropriate workplaces for this kind of learning.

Of course, optimum conditions for learning theories and main approaches through typical case studies are combined at university, before starting to work, and precise command of special methods, tools and gestures is naturally mastered during in the field practices. However, appropriate conditions for continual learning are not always available. It seems important to create appropriate context, to acquire just-in-time, missing knowledge.

For industrial firms, it could be appropriate to separate blended learning situations into alternative periods with theory and approach acquisition in a university context and their application and adaptation to precise industrial situations during associated industrial periods.

It is also possible to create individual conditions for this learning by e-learning, either out of working time (in the evening) or during working time (if allowed).

In the scope of this conference and our orientations, we are mainly interested in ICT enhanced learning. E-learning is the generic term for this field of studies as described in the introduction. As mentioned, we are specifically interested in Contextual Mobile Learning, the aim of which is to learn, just-in-time in the workplace, precise behaviors, actions and situations related to maintenance, diagnosis and repair of industrial appliances. For this kind of learning we identified three approach alternatives, but also the following complementary and cooperative approaches:

- Formal, firm-specific learning organization in direct relation with the firm's operational information,
- Web-based and forum-based access to world-wide accessible knowledge,
- Coupling between off-line learning of industrial practices by serious games and concrete working practical situations.

### A. Formal Approach EPSS

EPSS: Electronic Performance Support System is an appropriate answer to the problem of industrial in the field learning, which is formal, firm-specific, learning organization in direct relation with the firm's operational information [12]. An EPSS is a web-based content management system for

- Providing storage and delivery of plant reference materials including:
  ✧ Training documents
  ✧ Operating procedures
  ✧ Historical maintenance information
- Improving plant performance through knowledge transfer
- Storing knowledge captured from seasoned employees
- Providing procedure management

By using an EPSS, employees will receive the basic and additional support they need during their everyday tasks. The EPSS may be used as a training tool for new staff members or as a quick reference library for experienced personnel. The EPSS provides the user with information including information on plant systems and delivers plant procedures to the people that need them. The combination of digital photos, video, animation and sound is a powerful tool when attempting to stimulate a learner. All of these can be incorporated into the EPSS.

The aim of the EPSS is to improve human performance by providing staff with the tools they need to perform their jobs in a safe, efficient manner, which in turn, improves plant performance.

As presented later in this paper our approach is clearly compatible with this EPSS approach. However, it seems important to mention other more open-ended approaches.

*B. Open-ended approaches*

In their paper [13] "A Theory of Learning for the Mobile Age", Mike Sharples and his colleagues propose a new way to learn, based mainly on conversation. They propose a theory of learning for a mobile society in which new forms of educational activity take place based mainly on informal and workplace learning that is fundamentally mobile. This new learning is characterized as personalized, learner centered, situated, collaborative, ubiquitous and lifelong as related to new technologies characterized as personal, user centered, mobile, networked, ubiquitous and durable. In this approach that considers learning as conversation, teacher and learner jointly develop understanding through dialogue. In this way the teacher can progressively totally disappear and be replaced by the conversation between "partners" in a conversation web–based forum. These partners can be either humans or also machines, mainly in the form of a web accessible Knowledge base.

This practice is already observable in software development teams, where members of different and maybe rival teams can mutually help each other via a web-based forum to solve recurrent technological problems.

Another interesting point of view is expressed by Don Tapscott in his book "Grown up digital: How the net Generation is changing your world" [14]. He mentions new learner behavior as well as the behavior of workers, and also further interesting observations and expectations in relation with this new digital world.

*C. Mobile learning and Serious Game coupling*

Another interesting approach for industrial learning is the coupling between off-line learning of industrial practices by serious games and concrete working situations in the field. In this way it is possible to put into perspective the scenarios studied during the game and the concrete case on which the technician (learner) is working. With this connection he/ she can refresh the knowledge acquired during the game and re-contextualize it in the current precise situation. He/she can thus combine theories, methods and practices and thoroughly understand all interactions.

In our approach we take into account these aspects and we propose a contextualized mobile cooperative learning for industrial situations. In the following paragraphs we describe first different augmentations, as the way to create a continuum between the real and virtual (digital) world for contextualized working conditions for a human augmented technician (human) evolving in a real augmented environment and working on, observing, studying, diagnosing and repairing real augmented appliances.

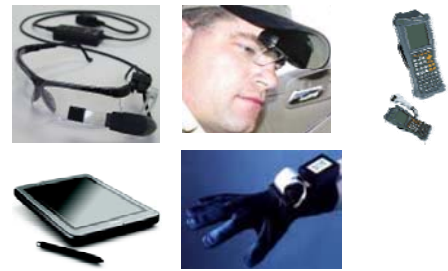

Figure 2. Goggles with integrated screen, See-Through goggles, TabletPC, RFID reader and Data glove

## III. AUGMENTATIONS

*A. Augmented human*

In order to create appropriate MOCOCO (mobile, contextual and cooperative) working and learning conditions, we provide the user (worker – learner) with a wearable computer and associated input/output devices (Fig. 2). Different users (actors) evolving in different working scenarios receive different configurations of wearable computers and associated peripherals. Several requirements can guide this choice: light and small hand free equipment, or heavier equipment but with better visualization capacities or better interaction performance …. For example, to present appropriate information to the user, screen integrating goggles or see-through goggles are carefully considered and chosen in order to create suitable Augmented Reality information allowing the user to concentrate on the repair work and associated learning. Elaboration of this configuration is based on a study of user's tasks, and their characterization by requirements concerning graphic information complexity (textual, graphic diagrams or precise blueprints, etc.), interaction complexity (writing, observation, manipulation) and working conditions (seating, standing, hands availability, etc.) and so on. A precise selection process based on a selection space allows comparison of different interaction modes and system implementations, with their typical supporting devices organized onto axes and classified for each axis by one of their most relevant characteristics [15].

*B. Augmented environment*

Our user (worker – learner) is evolving in a real augmented environment, which is able to provide him (by his/ her wearable computer and peripherals) with appropriate information mainly related to the context. In this way he/ she can decide what is relevant to do. The delivering of appropriate context information allows him/ her to decide what to do. This is a crucial aspect in mobile learning system design the aim of which is to deliver the right learning content into the right learning context.

In our industrial scenarios, we mainly use RFID and other associated technologies to collect contextual information. Radio-frequency identification (RFID) is an automatic identification method, relying on storing and remotely retrieving data using devices called RFID tags and RFID readers. The RFID tag can be applied to or incorporated into a product, animal, or person for the purpose of identification using radio-waves and can be read from several centimeters to meters away and beyond the line of sight of the reader. In our scenarios, each appliance registered in the manufacturer's database has a RFID tag, which indicates its own product serial number. The user equipped with an RFID reader on his/ her wearable computer is able to obtain immediately all the relative information and operation history about the appliance from the server. In this way, concerned learning units can receive precise context information which is used in the learning process to master or repair the appliance. We also explore the use of RFID tags and readers as part of the user interface and in relation with hand and body based interactions.

*1) RFID use*

In mobile context-aware applications, easy access to information describing and characterizing this context is important. Grounded communicating objects such as RFID tags can be used to deliver static information to the user's wearable computer RFID reader. This information can be used to access more precise and complete information which can be obtained from the EPSS system accessed by a wireless network. However, in situations in which wireless network is not systematically available and this information is mandatory, it seems important to be able to store these data in-situ.

*2) In-situ storage*

In situ storage is possible using RFID tags. These tags can be updated by the same reader, which is usually able not only to read but also to write information to the RFID tag. Storage capacity is extended significantly allowing storage not only of access identifiers, but of real information.

*3) Traceability*

In maintenance and repair of specific equipment traceability is a major requirement. The aim is to register all operations performed in order to be able to show these operations later, in order to verify their application or to understand or explain specific situations (behavior or accident). For this reason, the operations, tools used and product parts concerned can be collected and stored in a RFID tag or a central database (EPSS), tools and product parts are "tagged" by RFID tags and the user has a RFID reader in his wearable computer.

*4) Operation prescription*

Unlike traceability, execution of prescribed operations and respect of the operation sequence (workflow) can also be managed and controlled by RFID. To guarantee safety and reliability of the repairing process, it is possible to control respect of procedures. Control can concern actor identification, in order to determine his/ her accreditation level or the process, i.e. appropriate execution of the prescribed operation (sequence, tools and product parts).

*C. Augmented appliance*

In order to respond fully to HMTD problems, it is necessary to use wearable computers for different activities of diagnosis, maintenance, repair and associated M-learning. These computers must be correlated with technologies of the appliance on which the actor wants to act. Three forms of relationship between the wearable computer and the appliance can be identified (Fig. 3a, 3b and 3c):

- When the appliance does not propose any connection with the wearable computer, the user has to establish this connection indirectly by observing the appliance relating to the wearable computer observed situation and asking it to suggest an appropriate operation on the appliance (Fig. 3a);

- When the appliance is able to receive orders via, for example, an infra-red interface, it is possible to establish a unilateral contact from the wearable computer towards the appliance, to give it an order. The other method, which provides the wearable computer with the information observed on the appliance is replaced by user assistance (Fig. 3b);

- When the appliance is able to establish with the wearable computer two-directional communication, it is possible to substitute the original interface of the appliance for one proposed by the wearable computer interface. In this case this new interface can be totally adapted to user requirements (Fig. 3c).

*D. IMERA platform*

The IMERA platform designed to create an environment for experimentations consists of a main workplace and three auxiliary distant workspaces. The main working area is a CAE (Computer Augmented Environment) in which different mobile actors evolve. This CAE is an area of variable size covered by a WiFi network or another wireless network such as an adhoc network. The area is able to receive RFID tags, either freely set or integrated into real objects (augmented objects). The actors move freely in this area with their wearable computers, each equipped with a WiFi card and an RFID reader. These wearable computers are connected to the network and are able to collect contextual data via RFID technology. The WiFi network allows actors to be both connected with one another and with central systems (database servers, EPSS, etc.) so they can communicate and access large amounts of data. Independently from this working area, three separate distant management and observation workplaces complete this platform [11].

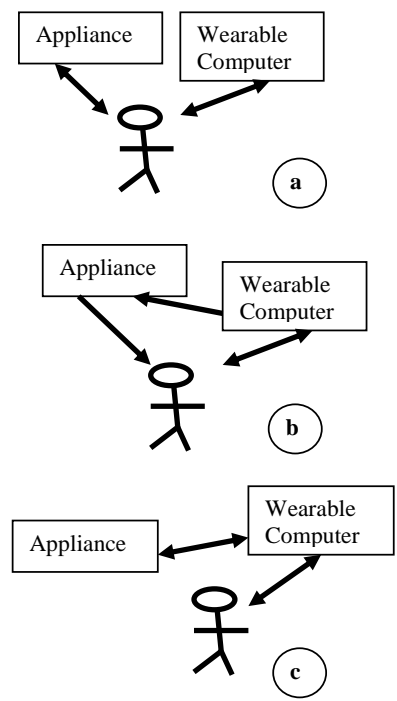

Figure 3. Appliance – Wearable computer communication relations

The IMERA platform can be used in a number of collaborative situations, managing educational, industrial, cultural and sporting events. For each application, it is important to identify the actors and their tasks with the data to be collected and manipulated. We thereby determine the technologies to use on the main working area and the most appropriate interaction devices for each actor. As we mentioned in the previous section, with the support of WiFi access points to internet, the user also has the possibility to collaborate with other actors located elsewhere. Communication can assume a variety of forms: voice, text message, drawing or gesture and so on. During the repair process of an appliance the technician can sometimes have problems checking out the reasons for dysfunction. He can ask for help and be guided in his/ her activity by his/ her technical supervisor. Other experts, anyone who can be located anywhere in the world can contribute to this collaborative problem solving.

## IV. GLOBAL PROCESS AND ORGANIZATION

### A. Learning unit production

Production of specific learning units for an appliance is schematized in Fig. 4 and is carried out by analysis of manufacturer documentation. For this step, an existing IMAT (Integrating Manuals and Training) project [16] provided us with inspiration in indexing and management of learning fragments. This, however, was initially designed for fixed expert users and did not consider mobile users and contextualization possibilities.

Information provided is not limited to the textual description of the tasks to be carried out (and thus learning), but can also contain diagrams, photographs, figures and models of elements (parts making up the equipment, its front face, its control panel, etc.). The goal is to build a computer supported model that is as complete as possible in order to be able to produce suitable learning units. The original source of documentation can assume a variety of forms, such as paper manuals, .doc, .pdf, video, voice, etc. Paper manuals are scanned and converted to electronic versions with OCR (Optical Character Recognition) technologies; other documents are also converted to uniform formats, which become learning fragments and finally learning units.

These learning units are segmented according to their usability, taking into account actors' experience characteristics i.e. beginner's actions, basic tasks, advanced tasks, expert tasks for mastering the use of appliances, as well as main tasks related to dysfunction identification, diagnosis process, repairing, dismounting the parts and reassembling, etc. The degree of specificity or generality of the explanation and later classification (indexing) is defined in order to be able to retrieve these units easily in the future. For specific industrial knowledge firm-protected units, they are stored in the EPSS, while for general knowledge units an open-ended web-based Knowledge base is used. All these units are expressed in XML using the standard SCORM and LOM to allow adaptation.

### B. IMERA working principles

During the working period, actors evolve in the IMERA workspace equipped with their wearable computers and associated peripherals, and work (use, master, diagnose and repair) on appliances (Fig. 5). The control engine of the actor's wearable computer assists him/ her with their activities and helps him/ her by providing appropriate working and learning units, in relation with the chosen method (working, just-in-time learning, learning and doing). It plays the role of a manager able to link concrete working situations with database stored learning units and is also responsible for adapting this information to visualization and interaction devices attached to the wearable computer.

Contextualization is provided by use of different markers, mainly RFID tags. In this way precise information on the appliance can be transmitted to the learning system. Actors (user, repairer) also use wearable computers to communicate with the system, between actors and, if possible, with the appliance.

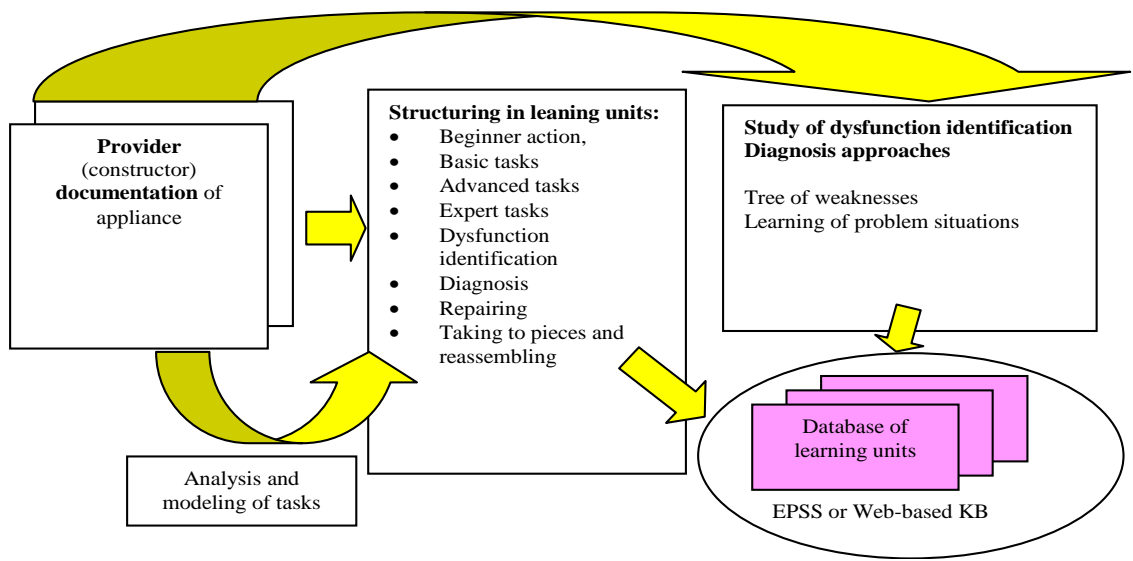

Figure 4. Learning unit production process

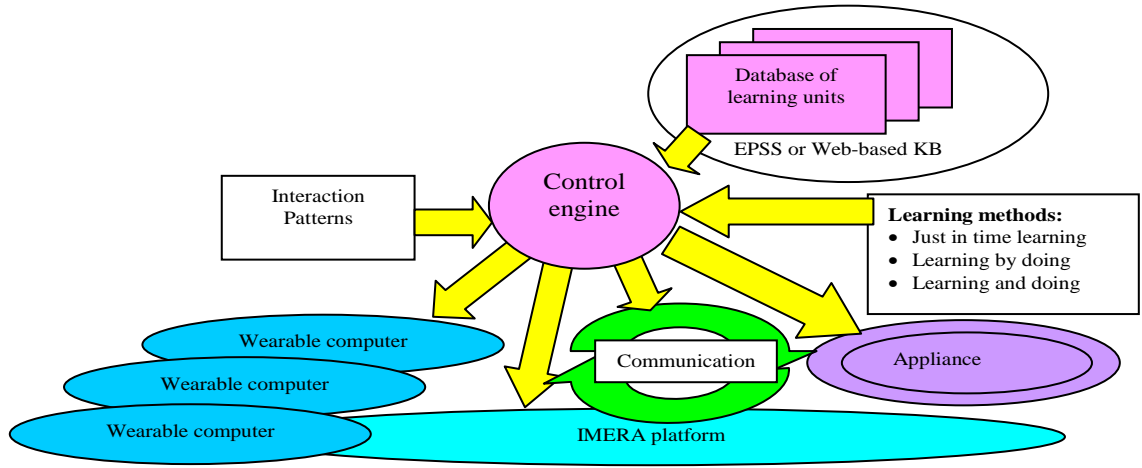

Figure 5. Learning units use on IMERA platform

## V. CASE STUDY

### A. Description of the case study

To explain the use of our MOCOCO system, we decided to present a commonly understandable problem, assistance with computer maintenance, and, more precisely, replacement of a hard disk of a desktop computer. We show how to assist a young technician who is carrying out, and learning about, replacement of a hard disk. Based on the contextual mobile learning principles described in the previous section, we propose a contextualized solution, using mobile devices (Tablet PC or PDA) and augmented reality accessories (semi-transparent glasses) to provide a just-in-time learning solution when the user needs a guide to do this task.

Through the contextualization process, the user immediately obtains brief information on the computer to maintain and a list of all possible operating guides on his/ her screen. This information helps him/ her to initiate the maintenance work.

In the scenario of changing the hard disk of a computer, work is made up of several steps such as: 1. Remove the screws on the case. 2. Remove the case. 3. Pull out the power connector…14. Fit on the case. If the user is not clear on one or more steps, he/ she can trigger the just-in-time learning process and learn how to carry out the actions.

He/ she can view the global list of actions or ask for more precise action by an explanation of each step activity. He/ she can watch the sequence of actions, listen to the voice guidance, etc… and perform his task step-by-step (Fig.6.).

If the user is equipped with augmented reality semi-transparent glasses, the system can show him/ her guided actions superimposed on real objects (Fig. 7).

If the user still experiences difficulties during work after individual learning, a collaborative interaction can launch communication with other remote experts via the network, to guide the user by appropriate methods (voice, text message, real gestures, etc.).

During this learning process, the user can not only finish the job by following the guide, but also try to understand the whole sequence of actions. Moreover, some additional information is displayed for the user to learn more about the task. For example, when the user removes the hard disk, he/she is informed on the screen of all the characteristics concerning the disk (Fig.6).

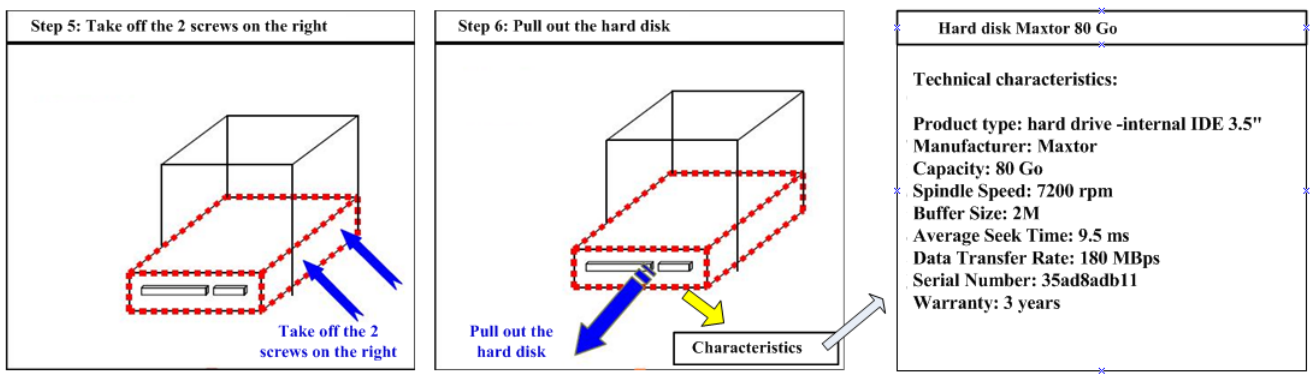

Figure 6. Changing hard disk manipulation process

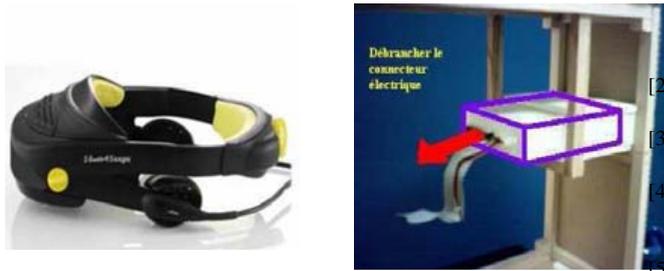

Figure 7. Semi transparent *see-through* augmented reality HMD

### B. Learning and applying which relationships?

We can propose three ways in which learning can be integrated in professional activities:

- Before work i.e. to learn about coming actions
- After work i.e. to learn about past actions to understand what happened and accumulate experience
- During work i.e. to master the problem just-in-time.

In our case study we can offer learners better mastery of different hard disk transfer rates, interfaces, multi device connections and so on, as explained in the appendix. Of course learners can either apply the method before, after or during their work and learn in identical manner, or apply only without in-depth understanding of what they have done.

## VI. CONCLUSIONS

M-learning (mobile learning) can take various forms. We are interested by contextualized M-learning, i.e. the training related to the situation physically or logically localized. The contextualization and pervasivity are important aspects of our approach. We propose in particular MOCOCO principles (Mobility - COntextualisation - COoperation) using IMERA platform (Mobile Interaction in the Augmented Real Environment) covering our university campus in which we are prototyping and testing our approach. From an organizational point of view we are in perfect symbiosis with EPSS - Electronic Performance Support System [12] and our objective is to integrate learning in professional activities in three ways: 1/ before work i.e. to learn about coming actions, 2/ after work i.e. to learn about past actions to understand what happened and accumulate experience, 3/ during work i.e. to master the problem just-in-time We are also studying an appropriate relationship between the just-in-time M-learning approach and preliminary training performed in a serious game approach based on typical action scenarios.

Different experimentations and evaluations of this approach and platform have been carried out in our laboratory, mainly with students as actors. The results are very positive and many useful suggestions were given. We are open to other applications to validate our approach.

We are also studying an appropriate relationship between the just-in-time M-learning approach and preliminary training performed in a serious game approach based on a typical action scenario.

## VII. APPENDIX

We show the nature of information which can be useful for the actor in charge of hard disk replacement. We list a set of knowledge items growing understanding of connection technologies. This information is coming from in Wikipedia : http://en.wikipedia.org/.

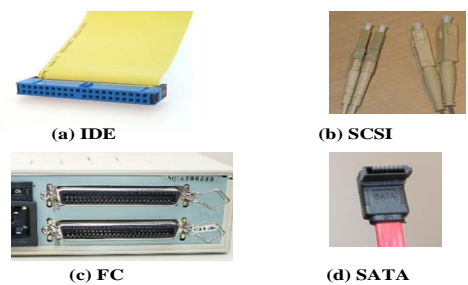

(a) IDE  (b) SCSI
(c) FC  (d) SATA

Figure 8. Main connection technologies

*A. Data transfer rate*

In 2008, a typical 7200rpm desktop hard drive has a sustained "disk-to-buffer" data transfer rate of about 70 megabytes per second. This rate depends on the track location, so it will be highest for data on the outer tracks (where there are more data sectors) and lower toward the inner tracks (where there are fewer data sectors); and is generally somewhat higher for 10,000rpm drives. A current widely-used standard for the "buffer-to-computer" interface is 3.0 Gbit/s SATA, which can send about 300 megabyte/s. from the buffer to the computer, and thus is still comfortably ahead of today's disk-to-buffer transfer rates. Data transfer rate (read/write) can be measured by writing a large file to disk using special file generator tools, then reading back the file. Transfer rate can be influenced by the fragmentation of the drive and the layout of the files. For example, a single file of 10GB will be read significant faster than 1000 files of 10MB.

*B. Interfaces*

Word serial interfaces: connect to a host bus adapter with two cables, one for data/control and one for power. The earliest versions of these interfaces typically had a 16 bit parallel data transfer to/from the drive and there are 8 and 32 bit variants. Modern versions have serial data transfer. The word nature of data transfer makes the design of a host bus adapter significantly simpler than that of the precursor HDD controller:

- IDE or ATA, P-ATA
- EIDE
- SCSI

Bit serial interfaces: connect to a host bus adapter with two cables, one for data/control and one for power:

- Fibre Channel (FC),
- Serial ATA (SATA).
- Serial Attached SCSI (SAS)

*C. Integrated Drive Electronics (IDE)*

Later renamed to ATA, and then later to P-ATA ("parallel ATA", to distinguish it from the new Serial ATA). The original name reflected the innovative integration of HDD controller with HDD itself, which was not found in earlier disks. Moving the HDD controller from the interface card to the disk drive helped to standardize interfaces, and to reduce the cost and complexity. The 40 pin IDE/ATA connection of PATA transfers 16 bits of data at a time on the data cable. The data cable was originally 40 conductors, but later higher speed requirements for data transfer to and from the hard drive led to an "ultra DMA" mode, known as UDMA. Progressively faster versions of this standard ultimately added the requirement for an 80 conductor variant of the same cable; where half of the conductors provide grounding necessary for enhanced high-speed signal quality by reducing cross talk. The interface for 80 conductors only has 39 pins, the missing pin acting as a key to prevent incorrect insertion of the connector to an incompatible socket, a common cause of disk and controller damage.

*D. Multiple devices on a cable:*

If two devices attach to a single cable, one must be designated as device 0 (commonly referred to as master) and the other as device 1 (slave). This distinction is necessary to allow both drives to share the cable without conflict. The mode that a drive must use is often set by a jumper setting on the drive itself, which must be manually set to master or slave. If there is a single device on a cable, it should be configured as master. However, some hard drives have a special setting called single for this configuration (Western Digital, in particular). Also, depending on the hardware and software available, a single drive on a cable can work reliably even though configured as the slave drive (this configuration is most often seen when a CD ROM has a channel to itself).

*E. EIDE*

It was an unofficial update (by Western Digital) to the original IDE standard, with the key improvement being the use of direct memory access (DMA) to transfer data between the disk and the computer without the involvement of the CPU, an improvement later adopted by the official ATA standards. By directly transferring data between memory and disk, DMA eliminates the need for the CPU and operating system to copy byte per byte. And can therefore process other tasks while the data transfer occurs.

*F. Small Computer System Interface (SCSI)*

It was an early competitor of ESDI. SCSI disks were standard on servers, workstations, Commodore Amiga and Apple Macintosh computers through the mid-90s, by which time most models had been transitioned to IDE (and later, SATA) family disks. Only in 2005 did the capacity of SCSI disks fall behind IDE disk technology, though the highest-performance disks are still available in SCSI and Fiber Channel only. The length limitations of the data cable allows for external SCSI devices. Originally SCSI data cables used single ended data transmission, but server class SCSI could use differential transmission, either low voltage differential (LVD) or high voltage differential (HVD).

*G. Fibre Channel (FC)*

It is a successor to parallel SCSI interface on enterprise market. It is a serial protocol. In disk drives usually the Fiber Channel Arbitrated Loop (FC-AL) connection topology is used. FC has much broader usage than mere disk interfaces, it is the cornerstone of storage area networks (SANs). Recently other protocols for this field, like iSCSI and ATA over Ethernet have been developed as well. Confusingly, drives usually use copper twisted-pair cables for Fiber Channel, not fiber optics. The latter are traditionally reserved for larger devices, such as servers or disk array controllers.

*H. Serial ATA (SATA)*

The SATA data cable has one data pair for differential transmission of data to the device, and one pair for differential receiving from the device, just like EIA-422. That requires that data be transmitted serially.

*I. Topology:*

SATA uses a point-to-point architecture. The connection between the controller and the storage device is direct. Modern PC systems usually have a SATA controller on the motherboard, or installed in a PCI or PCI Express slot. Some SATA controllers have multiple

SATA ports and can be connected to multiple storage devices. There are also port expanders or multipliers which allow multiple storage devices to be connected to a single SATA controller port.

*J. Serial Attached SCSI (SAS*

The SAS is a new generation serial communication protocol for devices designed to allow for much higher speed data transfers and is compatible with SATA. SAS uses a mechanically identical data and power connector to standard 3.5" SATA1/SATA2 HDDs, and many server-oriented SAS RAID controllers are also capable of addressing SATA hard drives. SAS uses serial communication instead of the parallel method found in traditional SCSI devices but still uses SCSI commands.

AUTHORS

**Bertrand DAVID, Chuantao YIN and René CHALON** are with the LIESP Laboratory, Ecole Centrale de Lyon, 36, Avenue Guy de Collongue, 69134 Ecully Cedex, France. (e-mail: {Bertrand.David, Chuantao.Yin, Rene.Chalon}@ec-lyon.fr).